# Art History on Wikipedia, a Macroscopic Observation

**Doron Goldfarb**
Vienna University of Technology
Favoritenstrasse 9-11
Vienna, 1040, Austria
doron@ec.tuwien.ac.at

**Max Arends**
Vienna University of Technology
Favoritenstrasse 9-11
Vienna, 1040, Austria
max@ec.tuwien.ac.at

**Josef Froschauer**
Vienna University of Technology
Favoritenstrasse 9-11
Vienna, 1040, Austria
josef@ec.tuwien.ac.at

**Dieter Merkl**
Vienna University of Technology
Favoritenstrasse 9-11
Vienna, 1040, Austria
dieter@ec.tuwien.ac.at

**Abstract**
How are articles about art historical actors interlinked within Wikipedia? Lead by this question, we seek an overview on the link structure of a domain specific subset of Wikipedia articles. We use an established domain-specific person name authority, the Getty Union List of Artist Names (ULAN), in order to externally identify relevant actors. Besides containing consistent biographical person data, this database also provides associative relationships between its person records, serving as a reference link structure for comparison. As a first step, we use mappings between the ULAN and English Dbpedia provided by the Virtual Internet Authority File (VIAF). This way, we are able to identify 18,002 relevant person articles. Examining the link structure between these resources reveals interesting insight about the high level structure of art historical knowledge as it is represented on Wikipedia.

**ACM Classification Keywords**
J.5 Arts and Humanities: Fine arts
H.3.1 Content Analysis and Indexing: Thesauruses
H.3.4 Systems and Software: Semantic Web

**Introduction**

While the structure of Wikipedia as a whole has been extensively studied, more focused observations of domain specific subsets of its content are often limited to qualitative comparisons on an individual article level, or on the extraction/analysis of underlying category schemes or taxonomies. With a growing crowd actively participating in the Wikipedia project, the amount of articles tied to specific knowledge domains has reached an extent that makes it feasible to observe their large-scale structure within defined domain borders.

Considering the domain of Cultural Heritage, a number of institutions are increasingly becoming aware of Wikipedia as a valuable tool to provide context for museum artifacts [4]. Indeed, the number of art history related resources on Wikipedia is impressive, e.g., searching only for distinct resources belonging to the category *yago:Painter110391653*[1] already returns 14,138 articles as of February 2012. There are, however, lots of different types of Wikipedia articles dealing with art history. There are articles about famous artworks (e.g. Mona Lisa), about artistic styles (e.g. Renaissance), about museums (e.g. Louvre) or about art historical actors (e.g. Leonardo da Vinci). Especially articles of the latter type are suited to a large-scale structural observation, as biographies of art history related people are highly interlinked due to student/teacher, artist/patron or influential relations.

In this study, we therefore focus on Wikipedia articles about art history related persons. We want to examine how they are related to each other and if there are interesting structural properties or groupings of inter-article links, such as if they reflect common attributes. As a first step, we want to approach the analysis by using visualization techniques.

But who are art historical actors? While it may be clear that a significant proportion are (groups of) artists like painters, sculptors, architects etc., there are also many other persons that have played major roles in the domain, such as important scholars, collectors, patrons, politicians or monarchs. This variety of different person roles makes it difficult to identify relevant persons. We therefore make use of a well-established external authority control, the Getty Union List of Artist names (ULAN)[2], in order to find art history related person biographies in extracted Wikipedia articles on Dbpedia. Besides supporting identification, this vocabulary also offers consistent biographical attributes such as birth/death-date, role or nationality, which is not always the case with Dbpedia [1]. Moreover, the ULAN provides associative relationships between its person records, which serve as a reference for comparison with the inter-article links between Wikipedia articles.

We make use of another authority, the Virtual Internet Authority File (VIAF)[3], for mapping ULAN records to Dbpedia resources. As of February 2012, there exist 18,002 of such mappings in VIAF. Since we use the Resource Description Framework (RDF) for the mapping between ULAN, VIAF and Dbpedia, this project also serves as an example on how Semantic Web technologies can be effectively used for combining different data sources, enhancing the quality of the data corpus and thus leading to interesting results.

---

[1] http://www.mpi-inf.mpg.de/yago-naga/yago/

[2] http://www.getty.edu/research/tools/vocabularies/ulan

[3] http://www.oclc.org/research/activities/viaf

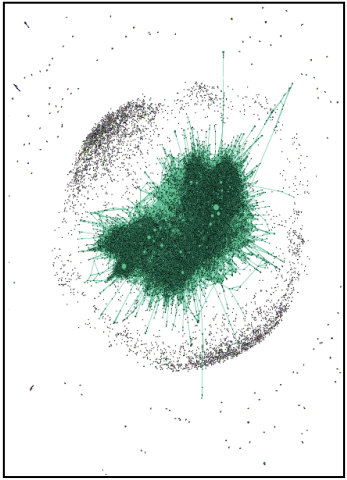

**Figure 1**: Visualization of 18,002 Dbpedia resources

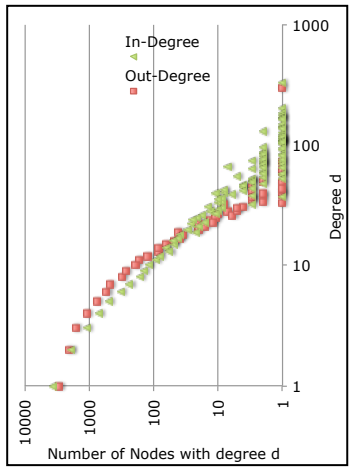

**Figure 2**: Power Law distribution of domain specific Wiki links

### Related Work

The large-scale structure of Wikipedia has been described in [6], pointing out that Wikipedia inter-article links form a scale-free network. A study focused on Wikipedia link structure is presented in [2]. The suitability of Wikipedia for deriving domain-specific thesauri is demonstrated in [3] by comparing its coverage with a professional thesaurus in the domain of agriculture. In [5], person related Wikipedia articles are analyzed for the domains philosophy and literature, with articles chosen by the people's occupation. The gathered information is used to show for the two domains which countries/cities have been important centers during the centuries. A comparison between ULAN and Dbpedia is presented in [1]. The authors were able to match 14,972 Dbpedia resources to their ULAN counterparts. Regarding Dbpedia coverage of biographical attributes, they were able to identify 8,077 (53.95%) attributes of type birth date, 9,628 (64.31%) of type nationality and 5,442 (36.35%) of type role. These findings encouraged us to rely on the ULAN as source for biographical attributes for the analysis of the link structure of the observed Wikipedia articles.

### Sytem Setup

We use RDF as data format, as it is the native format of the Dbpedia dataset, representing the core data model underlying the Semantic Web Vision. Like Dbpedia, we use Openlink Virtuoso[4] as storage platform. Using a tool provided by the MultimediaN N9C Eculture[5] project, we first converted the ULAN dataset to RDF. As the ULAN identifiers are by now completely integrated into VIAF, we were able to retrieve the VIAF RDF data for each of the ULAN records. Using the provided mappings, we downloaded the 18,002 RDF datasets from the public English Dbpedia Sparql endpoint. As this endpoint currently does not provide the internal Wikipedia links, we downloaded the complete internal wiki link dump from the Dbpedia website. We extracted all relevant wiki links between the 18,002 entities and loaded them into our triple store.

### A First Look at the Data

After gathering all the necessary data sources, we had a first look at the resulting structure by making a visualization using the Gephi[6] Platform. Figure 1 shows a force based layout of the network of all 18,002 Dbpedia resources and their 54,590 mutual links, where a node's size reflect its degree (in&out). The blob in the center suggests the existence of a giant connected component. Indeed, treating all links as undirected yields 4,187 different weakly connected components (CC), with a giant component - colored green in Fig. 1 - consisting of 75.44% (~13,581) of all articles, the rest being distributed across the remaining 4,186 CCs whose sizes range between 0.06% (~11) and single unconnected articles. Treating all links as directed yields 10,071 strongly connected components (SCC), with a giant SCC consisting of 40.55% (~7,300) of all articles, the remaining 10,070 SCCs counting between 0.05% (~9) and single unconnected articles. Looking at the node degree distribution, the link structure of the selected subset of articles clearly shows a Power Law distribution, as it has been identified with Wikipedia in general [6] - see Fig. 2. Observing the distribution of node in- and out-degree, we find comparable values as in [2].

---

[4] http://virtuoso.openlinksw.com

[5] http://e-culture.multimedian.nl

[6] www.gephi.org

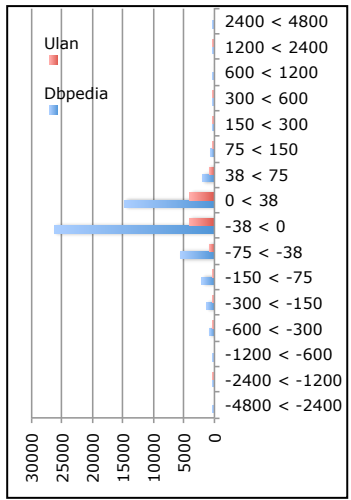

**Figure 3**: Distribution of link distances (linear scale)

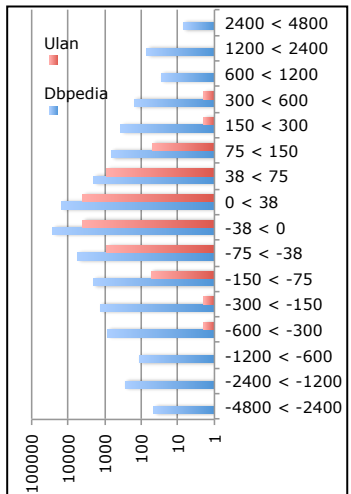

**Figure 4**: Distribution of link distances (logarithmic scale)

### Temporal Person-Person Link Properties

The network visualization in Fig. 1 is not very expressive, as the giant connected component visually resembles an unstructured, but highly connected "blob" of nodes. Therefore, we considered a closer look at the link properties, having a certain assumption in mind: As we observe only person articles, we can expect their mutual links to bear strong temporal characteristics, i.e., the timespan that a link covers ranges from zero years up to millenniums for persons referencing people from, e.g., ancient Greece. We approximate the interval for each person-person link by the difference between their birth dates as provided by the ULAN. Fig. 3 shows the distribution of link timespans on a linear scale for all the 54,590 Wiki links compared to all the 9,406 ULAN links for the same subset. What comes to attention first is that the number of "short" links covering 0-37.5 years clearly outnumbers the sum of all the other (Dbpedia: 75.3%, ULAN: 81.2%). This can be interpreted as such that contemporaries are much more likely to be interlinked than persons who are generations apart. We measure the link distance in both possible directions: into the future and into the past. This reveals an interesting difference between the associative links from ULAN and those from Wikipedia: While ULAN links are highly reciprocal, i.e., for almost every link from person A to B, there is also one from B to A, this is not the case in Dbpedia, where twice as many links point into the past (66.8%) than into the future (33.2%). Of the 54,590 Wiki links, 15,606 (28.59%) are mirrored and the remaining 38,984 are "unique". From a historical viewpoint, this seems quite logical, as artists can always draw from influences from previous generations, which are asymmetric by nature. From an encyclopedic viewpoint, however, this is not the case: While the Getty editorial guidelines for the inclusion of associative relationships in ULAN[7] (chapter 3.5.2.3.1) clearly state that each link has to be mirrored, this is not the case with the Wikipedia link style guidelines[8]. Nevertheless, the observed amount of mirrored Wiki links still differs significantly from [2], reporting about 8.7% reciprocal links for a much larger cross-domain subset of Wikipedia. This, however, might likely be related to our domain-specific set of person articles. Another difference between ULAN and Dbpedia links is that the latter do generally cover much larger timespans than the former - see Fig. 4 with link distances plotted against a log scale. This might also be explained through the ULAN editorial guidelines (chapter 3.5.1.4), stating that *"Relationships should be made only between records that are directly related"*.

### Refining the Visualization

The temporal properties of the Dbpedia links indicate that we might see more structure in the visualization by omitting links that extend beyond a certain time interval. As we use a force-based algorithm for the layout, we expect the node link structure to "unfold" into an approximately chronological sequence. Fig. 5 shows the result of applying the same layout algorithm as for Fig. 1, but without links whose endpoints are more than 75 years apart. We chose this distance, as 89.13% of all Dbpedia links are "shorter" (ULAN: 98.87%) and it is roughly suited to separate contemporary from inter-generational relationships. The resulting layout now has a chronological alignment, representing the flow of art history from the Renaissance (left) until today (right), solely based on the relations between its numerous historical actors.

---

[7] http://www.getty.edu/research/tools/vocabularies/guidelines

[8] http://en.wikipedia.org/wiki/Wikipedia:Manual_of_Style/Linking

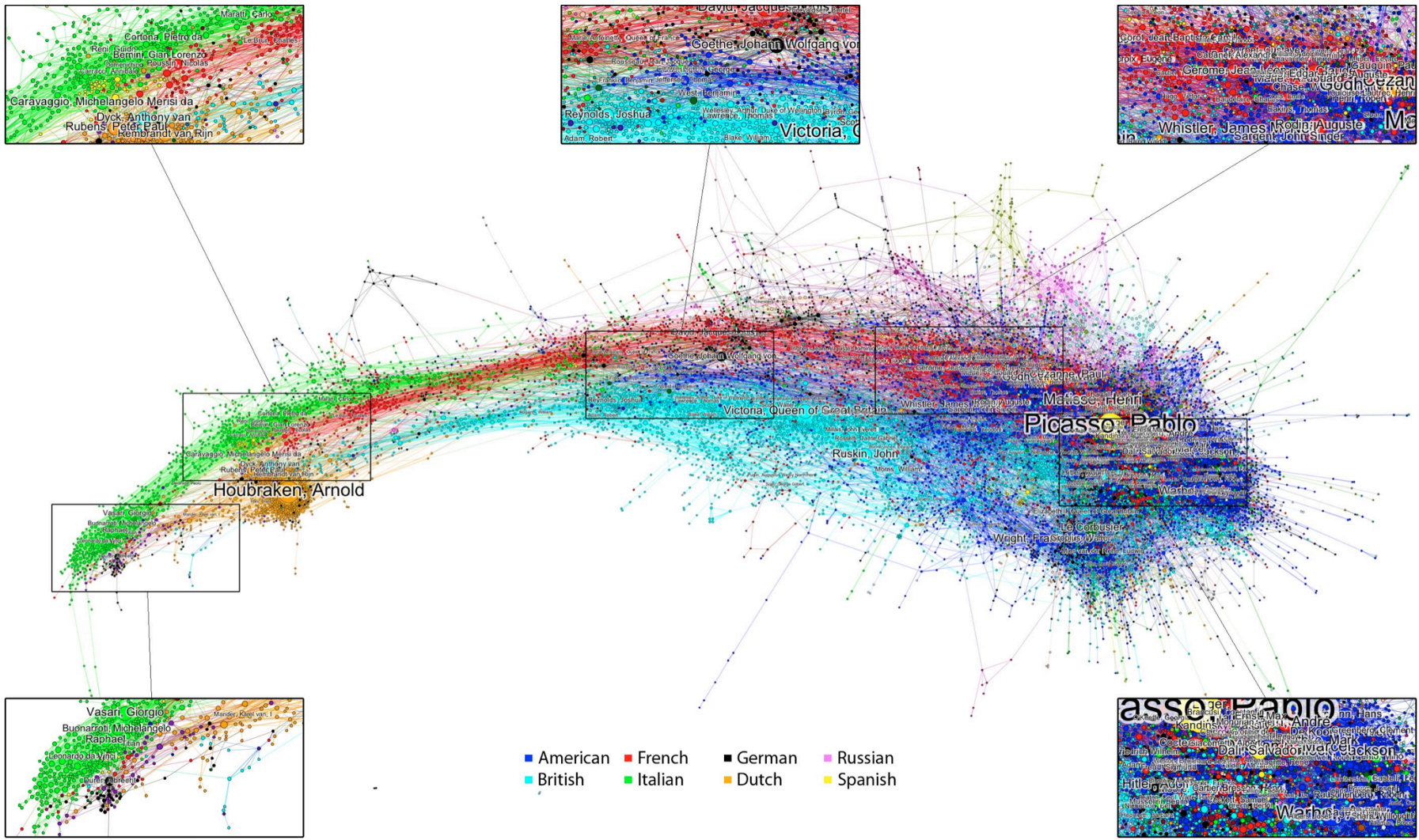

**Figure 5**: Visualization of Wiki links between art-historical actors with max. distance 75 years. A high-res version is available at http://www.wikiarthistory.info.

Coloring the nodes according to each person's nationality, as provided by the ULAN, reveals interesting patterns of cultural interaction within the network, as they are perceived by the English speaking Wikipedia community: The left side of Fig. 5 is dominated by Italians (green). This cluster spans Renaissance and Baroque times, fading out by the end of the 17th century. A small cluster on the lower left represents German Renaissance around Albrecht Duerer (black). Further right, a cluster of Dutch Renaissance and Baroque (orange) persons marks the beginning of an interesting development. Two distinct groups of British (cyan) and French (red) historical actors dominate the center, until Americans (blue) unite them into a big melting pot around the late 19th century. The rightmost part represents Post-Modernist Americans, with a nationality-independent cluster of Architects beneath.

**Discussion & Future Work**
By visualizing the reduced link set, we were able to identify structural characteristics of the inter-article links. There exist link based article clusters that represent the nationality of the respective persons and possess strong temporal characteristics that represent art historical periods and schools. A thorough discussion of these observations is, however, left for future work. As we have limited our observation to English Dbpedia and used a mainly Anglo-American authority file for the identification of articles, we expect to obtain different results for using authorities of different origin and for other language versions of Dbpedia. Comparing the results could reveal differences in cultural perception between the different Wikipedia communities. Furthermore, it would be interesting to apply the described method to other domains.

**Conclusion**
We have effectively integrated heterogeneous data sources for the analysis of 54,590 links between 18,002 domain specific Wikipedia person articles. Our findings include indications that properties such as node degree distribution are comparable to entire Wikipedia, while others like reciprocity of links seem different. The observed links tend to cover timespans comparable to a human lifetime, and are more likely to point into the past than vice-versa. By removing links covering more than 75 years, we visually identified article clusters with interesting cultural and temporal properties.

**Acknowledgements**
This work was funded by the FWF Austrian Science Fund, Project No. L602, *The Virtual 3D Social Experience Museum*